# Unusual field dependence of radio frequency magnetoimpedance in $La_{0.67}Ba_{0.33}MnO_3$

A. Rebello and R. Mahendiran

*Department of Physics and NUS Nanoscience & Nanotechnology Initiative (NUSNNI), Faculty of Science, National university of Singapore, 2 Science Drive 3, Singapore -117542*

Abstract

Magnetic field dependence of the *ac* impedance $Z(f, H) = R(f, H) + jX(f, H)$ in $La_{0.67}Ba_{0.33}MnO_3$ carried out over a wide frequency range ($f = 0$ to 30 MHz) reveals a huge low-field *ac* magnetoresistance, $\Delta R/R = -55\%$ at $f = 15$ MHz and magnetoreactance, $\Delta X/X = -80\%$ at $f = 3$ MHz in $\mu_0 H = 100$ mT at room temperature. We show contrasting evolution of $\Delta R/R$ and $\Delta X/X$ with increasing magnetic field and frequency. While $\Delta R/R$ is negative and shows a single peak at $\mu_0 H = 0$ T for all but $f = 30$ MHz, the single peak in $\Delta X/X$ transforms into a valley at the origin and a double peak at $H = \pm H_K$ which shifts upward in $H$ with increasing frequency. The $\Delta X/X$ eventually changes sign from negative to positive above 25 MHz. The observed features in $\Delta X/X$ suggest possible occurrence of ferromagnetic resonance in MHz range.

Perovskite manganese oxides of the general formula $R_{1-x}A_xMnO_3$ (R = trivalent rare earth ion, A = divalent alkaline earth ion) have been a topic of intensive investigation for the past one decade due to the spectacular negative magnetoresistance effect shown by them [1]. While several exotic electronic and magnetic phases exhibited by these oxides shifted recent interest in these materials towards more fundamental studies on strongly correlated electron systems [2], technological exploitation of the CMR effect at room temperature has been hampered by the need of magnetic field of $\mu_0 H > 1$ Tesla to induce more than 10 % magnetoresistance (MR). Extensive efforts are underway in different laboratories to enhance the magnitude of magnetoresistance at low fields, particularly in milli and micro tesla range using techniques such as artificial grain boundary in epitaxial films [3], trilayer tunnel junctions [4] and step edge junctions [5]. While magnetotransport published in the last decade, focused mostly on magnetoresistance measured with a direct current, high frequency electrical transport received less attention except for a few studies on microwave absorption ($f$ = 9-40 GHz) using cavity perturbation method [6-8] and radio frequency absorption using a transmitter-receiver coil [9,10] or tunnel diode techniques [11]. A substantial increase in *ac* magnetoresistance is reported by passing *rf* current directly though the sample [12] or through an inductance coil surrounding a manganite sample [13]. Recently, Barik et *al*. [14] has reported a spectacular increase in the electrical impedance of $La_{0.25}Sr_{0.75}MnO_3$ at the Curie temperature in the absence of external magnetic field and a large low field *ac* magnetoresistance (= 19 % for $\mu_0 H$ = 65 mT, $f$ = 3 MHz at $T$ = 300 K), by passing radio frequency current directly through the sample. However, a detail investigation of magnetic field dependence of both the components of electrical impedance is currently lacking.

In this work, we have investigated the low magnetic field response of the *ac* electrical impedance in $La_{0.67}Ba_{0.33}MnO_3$ at room temperature and show that *ac* magnetoresistance is extremely high ($\Delta R/R$ % = $[R(H)-R(0)]/R(0)$ *100 ≈ -55 % in $\mu_0 H$ = 100 mT for $f$ = 15 MHz) compared to a smaller *dc* magnetoresistance (< 2 % at $\mu_0 H$ = 100 mT) at the same field strength. The inductive magnetoreactance ($\Delta X/X$ % = $[X(H)-X(0)]/X(0)$) reaches as large as -80 % for $f$ = 0.1-3 MHz and decreases at higher and lower frequencies. A distinct evolution of magnetoresistance and magnetoreactance with increasing magnetic field and frequency are also found. While the *ac* magnetoresistance is negative and shows a single peak centered at the origin for all the frequencies, the magnetoreactance as a function of magnetic field shows a transition from a single to double peak structure with increasing frequency and eventually a change of sign.

We have measured the four probe *ac* impedance $(Z(\omega,H) = R(\omega,H) + j X(\omega,H))$ of polycrystalline $La_{0.67}Ba_{0.33}MnO_3$ rectangular bar of dimensions ≈ 11 mm x 3.4 mm x 1.7 mm with a SRS850 Lock-in amplifier for low frequency range ($f = \omega/2\pi$ = 30 Hz to 90 kHz ) and an Agilent 4285A LCR for high frequency range ($f$ = 75 kHz to 30 MHz) with 3 mA *ac* excitation current . The distance between voltage probes was fixed to 6 mm. An electromagnet provided a variable *dc* magnetic field ($\mu_0 H$ ) from -100 mT to +100 mT and the field was applied either parallel to the long axis of the sample, i.e., along the direction of the *ac* current (longitudinal configuration) or perpendicular to the long axis of the sample, i.e, perpendicular to the *ac* current (transverse configuration).

Figure 1 (a) shows the evolution of the *ac* resistance (R) under $\mu_0 H$ = 0 T and 100 mT while sweeping the frequency from $f$ = 30 Hz to 30 MHz. The $R (\mu_0 H = 0$ T$)$ increases nonlinearly as the frequency increases. A drastic decrease in the resistance occurs under the magnetic field for $f$ above 5 MHz. On the other hand, $X (\mu_0 H = 0$ T$)$ increases rather linearly up to 5 MHz, exhibits a plateau in the frequency range 10 – 20 MHz and then decreases at higher frequencies (see figure 1 (b)). The plateau is suppressed under the magnetic field and $X$ increases almost linearly with frequency up to 30 MHz. Interestingly, the $X$ ($\mu_0H$ = 100 mT) curve crosses the $X(\mu_0H$ = 0) curve around 25 MHz. The *ac* magnetoresistance *($\Delta R/R$ (%))* and the magnetoreactance, *($\Delta X/X$ (%))* are shown on the left and right scales, respectively in figure 1 (c). The magnitude of the *ac* magnetoresistance increases from less than -2 % at 30 Hz to a maximum of ≈ -58 % at 18 MHz and then decreases slightly to ≈ -50 % at 30 MHz. In contrast to the behavior of *$\Delta R/R$*, *$\Delta X/X$* shows a maximum (≈ -80 %) around 2 MHz with a sharp decrease on the low frequency side compared to the high frequency side. Eventually, *$\Delta X/X$* changes sign from negative to positive above 25 MHz.

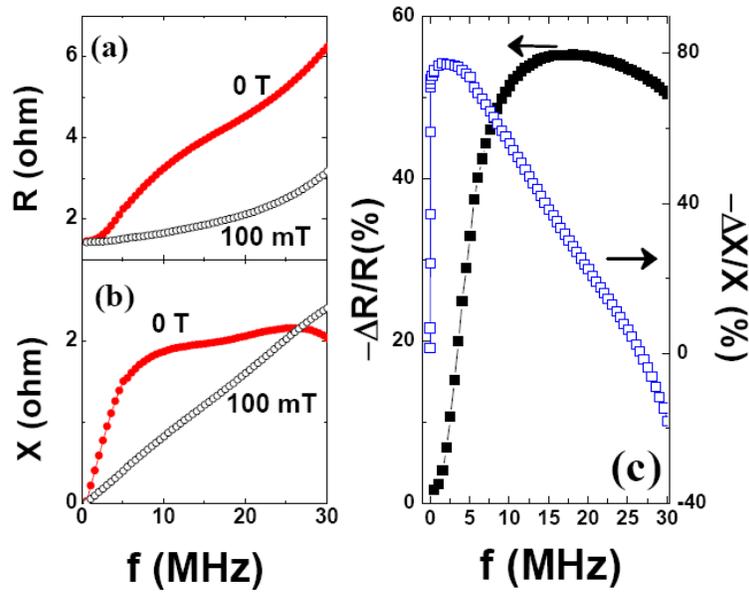

Figure 1: Frequency dependence of (a) the *ac* resistance, $R$ and (b) the inductive reactance, $X = \omega L$ in $\mu_0H$ = 0 T (solid symbol) and 100 mT (open symbols). (c). The *ac* magnetoresistance *($\Delta R/R$)* and the magnetoreactance *($\Delta X/X$)* as a function of frequency are shown on the left and right scales respectively.

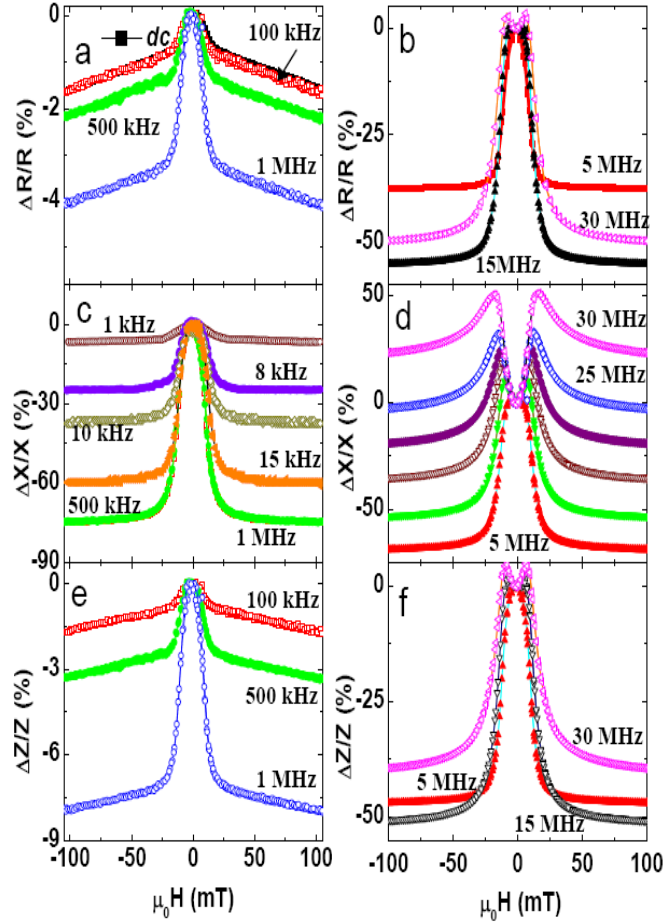

Figure 2: (Top panel) Magnetic field dependence of the *ac* magnetoresistance (*ΔR/R*) at room temperature for (a) $0 \leq f \leq 1$ MHz (b) $5 \leq f \leq 30$ MHz. The middle panel shows the magnetoreactance (*ΔX/X*) and the bottom panel shows the total magnetoimpedance (ΔZ/Z).

Figure 2 shows the magnetic field dependence of *ΔR/R*, *ΔX/X* and *ΔZ/Z* at 300 K for selected frequencies. The top panel shows longitudinal *ΔR/R* for (a) $0 \leq f \leq 1$ MHz and (b) $5 < f \leq 30$ MHz. We have shown the data for a few fixed frequencies though data at each 2 MHz interval are recorded. In the *dc* limit ($f = 0$ Hz), the maximum MR is $< -2$ % at $\mu_0 H = 100$ mT and it shows a marginal increase ($\approx -4$ %) when $f = 1$ MHz. However, *ΔR/R* increases to $\approx -37$ % at 5 MHz, attains a maximum value of $\approx -55$ % at 15 MHz and then decreases to $\approx -50$ % at 30 MHz in close agreement with the frequency sweep data in figure 1 (c). Our *ΔR/R* value is much larger than the earlier report by Hu *et al*. [12] who found a maximum 10 % at $f$ = 23 MHz in $\mu_0 H$ = 0.5 T in a similar composition. In contrast to the $\Delta R/R$, the magnitude of the magnetoreactance, $\Delta X/X$, increases rapidly from < -5 % at 1 kHz to -80 % at 100 kHz and remains unchanged even at 1 MHz (see figure 2(c)). The $\Delta X/X$ decreases in magnitude as the frequency increases above 5 MHz (see fig. 2(d)). Interesting field dependence is seen with increasing frequency. While the $\Delta X/X$ shows a single peak at the origin for $f \leq 8$ MHz, the single peak transforms into a double peak symmetrical positioned at $H = \pm H_k$ when $f$ = 10 MHz. As the frequency increases, $H_k$ shifts towards higher field and the amplitude of the double peak also increases. Eventually, $\Delta X/X$ changes sign from negative to positive in the entire field range at $f$ = 30 MHz. A double peak of much smaller amplitude is also visible in $\Delta R/R$ at 30 MHz (see the top right panel). The total low-field magnetoimpedance, $\Delta Z/Z$ is dominated by the *ac* MR and attains a maximum value of ≈ -55% between 10 and 15 MHz which is remarkably larger than the -2 % *dc* magnetoresistance observed (see figure 2(e) and figure 2(f)).

The observed features in $\Delta R/R$ and $\Delta X/X$ depend on the angle between the direction of the applied *dc* magnetic field and the *ac* current in the sample. We compare the magnetic field dependence of the longitudinal and transverse $\Delta R/R$ for (a) $f$ = 20 MHz and (b) $f$ = 30 MHz in figure 3. The corresponding data for $\Delta X/X$ are shown in the bottom panel for (c) $f$ = 20 MHz and (d) 30 MHz. The rapid change observed at low fields in the longitudinal magnetoresistance at $f$ = 20 MHz is absent in the transverse magnetoresistance which shows a gradual decrease above 55 mT. The transverse MR is lower in magnitude and it exhibits a pronounced hysteresis compared to the longitudinal case. A similar behavior is observed in $\Delta R/R$ at $f$ = 30 MHz. The difference between the two configurations is clearly reflected in the $\Delta X/X$. The double peak in the transverse configuration is suppressed in magnitude and its position is shifted to $\mu_0 H_k = \pm 75$ mT at $f$ = 20

MHz ($\mu_0 H_k > \pm 100$ mT at $f = 30$ MHz) compared to $\mu_0 H_k = \pm 14$ mT ($\mu_0 H_k = \pm 18$ mT at $f = 30$ MHz) in the longitudinal case. The transverse $\Delta X/X$ is positive for both the frequencies in contrast to the longitudinal case in which $\Delta X/X$ changes sign with increasing field.

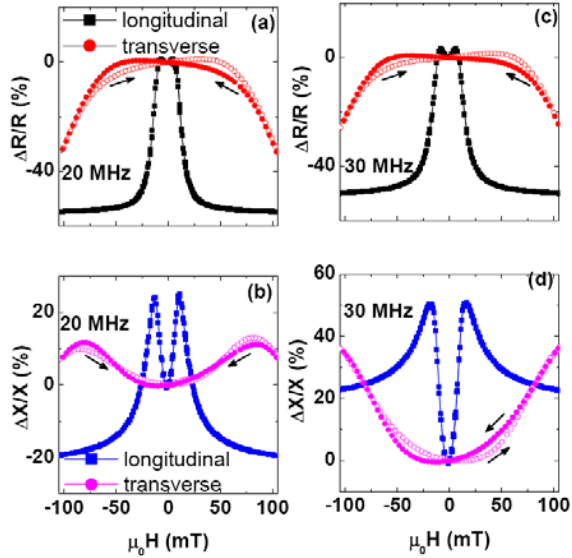

Figure 3: Magnetic field dependence of $\Delta R/R$ and $\Delta X/X$ in the longitudinal and transverse geometries for $f = 20$ MHz and 30 MHz. The arrows indicate the direction of field sweep.

What are origins of the huge low-field *ac* $\Delta R/R$ and double peak in $\Delta X/X$? In the *dc* limit ($f = 0$), the rapid decrease in resistance at low-fields is generally attributed to the spin-polarized electron tunneling between ferromagnetic grains separated by the thin semiconducting grain boundaries. We are unaware of any theoretical model which takes care of the frequency dependent tunneling in explaining the low field magnetoresistance in manganites. Though the huge *ac* MR observed below 25 mT and above 1 MHz can be attributed to the enhanced tunneling at higher frequencies, there can be other mechanism(s) operating at higher frequencies. Let us first consider the low frequency behavior. The low frequency *ac* current flows uniformly throughout the volume of the sample and the low frequency impedance can be written as $Z(\omega) =$ $R(\omega) + j\omega L(\omega)$. Since the frequency dependence of the inductance can be written as $L(\omega) = G\mu(\omega) = G(\mu'(\omega) - j\mu''(\omega))$ where $G$ is a geometrical factor, $\mu$ is the complex permeability, impedance becomes $Z = [R(\omega) + G\mu''(\omega)] + j\omega\mu'(\omega)$. Thus, the real part of the impedance represents not only the ohmic loss but also the magnetic absorption. The huge magnetoreactance observed below 1 MHz with a smaller enhancement in magnetoresistance can be easily traced to the decrease in $\mu'$.

However, as the frequency of the *ac* current increases, the flow of *ac* current is restricted to a layer of thickness, called skin depth or penetration depth $\delta$ ($\delta = \sqrt{(2\rho/\mu_0\mu_t\omega)}$) which leads to increase in the electrical impedance compared to the low frequency case. In our sample which has $\rho(300\text{ K}) = 56$ m$\Omega$ cm, the non magnetic skin depth (taking $\mu_t = 1$) are $\delta_0 = 11.95$ mm, 3.78 mm and 2.18 mm at $f = 1$, 10, and 30 MHz, respectively; and these values are larger than the thickness of our sample ($2t = 3$ mm) except at $f = 30$ MHz. However, the magnetic skin depth can decrease drastically by an order of magnitude if we assume $\mu_t = 100$. The large enhancement of $\Delta R/R$ between 1 MHz and 5 MHz possibly indicates that the magnetic skin depth is comparable to the thickness of the sample at least at 5MHz. Hence, it is reasonable to consider that $\delta < 2t$ above 5 MHz. In the high frequency limit, $Z = (1+j)\frac{\rho}{\delta} = \sqrt{j\omega\rho(H,\omega)\mu(H,\omega)}$. Since the transverse permeability is a complex quantity, the impedance becomes $Z = \sqrt{\omega\rho(H)}\left[\sqrt{\mu_R} + j\sqrt{\mu_L}\right]$, where $\mu_R = \sqrt{(\mu'^2 + \mu''^2)} + \mu''$ is the resistive loss and $\mu_L = \sqrt{(\mu'^2 + \mu''^2)} - \mu''$ is the inductive reactance. We have ignored dispersion in the resistivity since it is not expected to be significant for conduction electrons in the MHz range. The real part of the impedance represents power absorption by the sample [15] and it depends on the square root of the product of both the ohmic loss and the magnetic loss. Since both the transverse $\mu'$ and $\mu''$ can decrease drastically in a small magnetic field compared to the resistivity, $Z'$ shows a dramatic decrease with increasing field, particularly in low magnetic fields. Alternatively, it can also be said that the applied magnetic field reduces the transverse permeability which increases the skin depth and hence enhances the *ac* magnetoresistance in low magnetic fields.

Both domain wall oscillation and domain magnetization rotation contribute to the transverse permeability at low frequencies. At higher frequencies, the domain wall oscillation is damped out due to microeddy currents and hence magnetization reversal occurs predominantly via domain magnetization rotation. Suppose the sample is a single domain and the magnetization point along the easy axis determined by the anisotropy field ($H_k$), and the *dc* magnetic field ($H_{dc}$) is applied along the hard axis which is normal to the direction of $H_K$, then abrupt rotation of domain magnetization towards the *dc* field direction occurs when $H_{dc} = H_K$ [16]. As the *dc* magnetic field increases, the transverse susceptibility ($\chi_t = \mu_t - 1 = M_s/(H_{dc} - H_K)$, where $M_s$ is the saturation magnetization) is expected to diverge at $H_{dc} = H_K$. As a consequence, a peak occurs in the impedance at $H_{dc} = H_K$. Thus, the transition from single to double peak can be suggested to the suppression of the domain wall oscillation and the growing contribution of domain magnetization rotation to the transverse permeability.

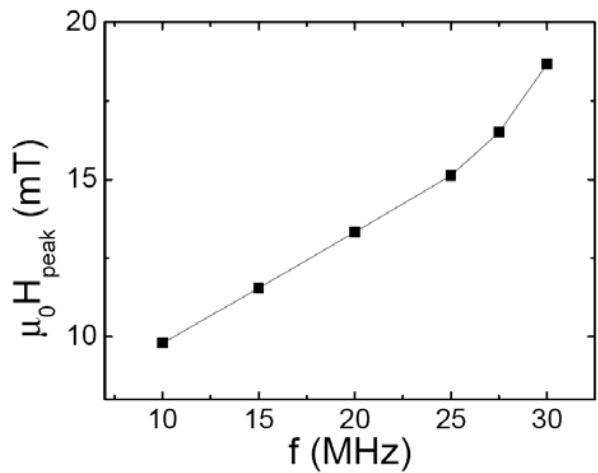

Figure 4: The shift of the anisotropy field ($H_k$) corresponding to the double peak in $\Delta X/X$ as a function of frequency.

The shift of $H_K$ towards a higher field with increasing frequency resembles the behavior of ferromagnetic resonance. The ferromagnetic resonance is characterized by a maximum in $\mu''$ or $Z'$, while $\mu'$ or $Z''$ crosses zero at a field corresponding to the ferromagnetic resonance [15]. Note that the inductive reactance ($\mu_L$) goes to zero if $\mu' = 0$ and so is the $Z''$ as we observed. In fact, the case of longitudinal magnetoimpedance in which smaller *ac* magnetic field and a larger *dc* magnetic field transverse to each other is exactly the geometry encountered in the ferromagnetic resonance experiments except that the *ac* field in the latter case is provided by an external *rf* coil. Though ferromagnetic resonance is generally observed in the GHz range in saturated ferromagnetic samples, it can also appear in the MHz range in multi domain (unsaturated state) state [17] or in the internal anisotropy field ($H_K$) in the sample. The ferromagnetic resonance frequency in a saturated sample is affected by the saturation magnetization, demagnetizing factor, anisotropy field, magnetoelastic strain and also by porosity in polycrystalline samples [18]. Recently, ferromagnetic resonance–like behavior in magnetoimpedance (a change in sign of $X$ and shift of $X$ with frequency) was reported in a few materials that include amorphous CoFeSiBNb wire around 900 kHz [19], NiFe/Cu/NiFe multilayers around 500 MHz [20], and Fe-Co-Si-B amorphous ribbon around 100 MHz [21]. The occurrence of both ferromagnetic resonance and antiferromagnetic resonance in magnetoimpedance was also demonstrated in certain amorphous ribbons [22]. However, signature of ferromagnetic resonance in the MHz range in manganites has not been reported so far. The ferromagnetic resonance frequency for a magnetically saturated sample is expected to shift with the applied *dc* magnetic field following the Kittel's relation $f_r^2 = (\gamma/2\pi)^2 \left[ H_0 + (N_y + N_y^a - N_z)M_s \right]\left[ H_0 + (N_x + N_x^a - N_z)M_s \right]$, where $\gamma$ is the gyromagnetic ratio, $H_0$ is the applied *dc* magnetic field, $N_x$, $N_y$, and $N_z$ are the demagnetization factors along the x, y, z axis, $N_y^k = H_y^k/M_s$ is the demagnetizing factor due to anisotropy field $H_k = 2K/M_s$, $K$ is the anisotropy constant and $M_s$ is the saturation magnetization [23]. If we approximate our sample as a long cylinder and the long axis of the sample along the z axis, then $N_x = N_y = ½$, $N_z = 0$, $N_y^k = N_x^k = 2K/M_s^2$, then the resonance frequency is given by $f_r = (\gamma/2\pi)(H_0 + M_s/2 + 2K/M_s)$. We have plotted the shift of $H_K$ with frequency in figure 4. The peak shifts almost linearly with magnetic field up to 20 MHz and a clear up turn is visible around 25 MHz. Because of the maximum frequency limit (30 MHz) in our experiment, we are unable to unambiguously attribute the observed effect to the ferromagnetic resonance though it appears to be a plausible origin. Another possibility is the magnetoelastic resonance of domain walls. Maartense and Searle [24] found transition from single to double peak behavior in the *rf* transverse susceptibility of α-$Fe_2O_3$ below 80 MHz and attributed to the coupling between crystal's acoustic resonance modes and low-lying spin-wave modes excited near local crystal strains. Recently, *rf* current induced domain wall resonance was also reported in permalloy nanowire [25].

In summary, our investigation of radio frequency ($f$ = 1-30 MHz) magnetoimpedance study in $La_{0.67}Ba_{0.33}MnO_3$ reveals a huge *ac* magnetoresistance (= -55 % in $\mu_0H$ = 100 mT, $f$ = 10 MHz) compared to the smaller *d*c magnetoresistance (< -2 %) at room temperature. It is shown while the magnetoresistance and magnetoreactance exhibit a singe peak centered at $\mu_0H$ = 0 T for $f \leq 8$ MHz, the single peak in magnetoreactance transforms into a valley at the origin and a double peak at $H = \pm H_k$ which shifts upward in field with increasing frequency. While the magnetoresistance is negative, the sign of magnetoreactance changes from negative to positive at the highest frequency. It has been suggested that the high frequency effect is possibly related to the occurrence of ferromagnetic resonance. Further studies above 30 MHz and in a completely saturated state will be useful to understand the origin of the observed frequency dependent shift in

$\Delta X/X$.